# On migration to Perpetual Enterprise System

*M.T. Carrasco Benitez - dragoman.org/pes - 8 October 2023*

**Overview**

This document describes a pragmatic approach on how to migrate enterprise computer systems to new systems that could evolve forever, address the whole organisations and that are integrated.

Governance aspects are as important, if not more, than purely technical IT aspects: human resources, supply chains, call for tenders, and similar. Migration implies not starting from a green field.

**Style of this document**

{Principle} *Lie* if it helps and *restate* the obvious.

Enterprise IT architecture is a complex field. Efforts have been made to make this document accessible to the widest possible public, including non-IT people. To make concepts more accessible, they might be introduced informally without being technically strict (*lie*) and sprinkled with bits of tutorials (*restate*). For the gory details follow the references.

**Current systems**

{Definition} *Current system*.- The set of all IT artefacts in an organisation. It could be anything: from one integrated system to many disconnected systems, from properly developed systems to spreadsheets, from internal developed code to external libraries, etc.

The first priority is to ensure the functioning of the current system, imperfect as it might be. Avoid the syndrome of not maintaining the current system because it is a waste of money. It is an error to channel most of the IT resources into the new wonderful system on the way.

The first step is to prepare *emergency manuals* for the current system. The guiding scenario for preparing these manuals is that present IT staff operating/maintaining the current system disappear from one day to the next; unpolished manuals would do. New IT replacement staff without any previous knowledge should have a sporting chance of operating/maintaining the current system with the help of emergency manuals which must be printed and stored in a place easy to find.

Current systems can be viewed as inhabited towns. One cannot empty a town of its inhabitants (where do they go in the mean time?), flatten it and rebuild it from scratch starting with a clean sheet of paper. Inhabitants must continue living in their imperfect town which has to evolve addressing requirements in order of priorities [R1958].

**Wrong approach**

{Definition} *Legacy system*.- System that cannot be properly operated and maintained due to any types of shortcomings: human expertise, documentation, hardware replacement, software infrastructure maintenance, external services, sufficient budget, and legal certainty. Corollary 1: a bad system is a bad system, not a legacy system. Corollary 2: bespoke systems become legacy systems very quickly.

In other words, **legacy system is largely a human resources and supply chain issue**; it is not a technical IT issue [TBC].

A typical organisation has a mixture of unintegrated computer systems: legacy systems with code that nobody understand anymore and dare to touch, the original authors long gone [OC]; proprietary systems at the end of supported life; some systems with both of these vices and a few more vices thrown in for good measure [UKB].

{Definition} *Organisation IT history*.- Description of all the organisation IT systems, subsystem, programs or components: present or past, including unfinished and/or never commissioned systems; internal and external development and/or hosting, including as a service system; relation among the systems; responsible parties, department and/or person; costs; deep or superficial description, according to the available information, even just knowing the systems names. Abbreviate to *IThistory*.

Organisations tend to have a pronounced IT amnesia. Completing the IThistory might represent looking to the previous fifty years. Detailed documents with requirements, costs and similar details might be unavailable. Annual plans and activity reports usually include overview of computer projects and they would reveal at least the existence of projects. Some information might be just in the memories of people: one should talk to staff with memories of previous projects, it might also be very revealing talking to long gone former staff, perhaps retired. This is not a costly exercise and at the very least, it should help repeating some previous mistakes. A historian with IT know-how is the best person to create the IThistory.

The *proper system syndrome* is common: develop a new beautiful system following all the latest fads and theories that will cover all requirement in a magic way; IThistory is at least a partial cure for this syndrome. A typical course of action is to commission a study from a reputable consulting firm that will deliver a thick (physical and conceptual) specification, ask another external firm to build the new system, and after a few years install the wonderful new system in a big-bang: **it never works**.

{Principle} Abolish external pseudo-consultants [MC][CAU].

Often, large external consulting firms are more a hindrance than a help, and a very costly one. The inexpert supplied pseudo-consultants tend to have limited real technical knowledge, in addition to no knowledge about the organisation [TBC].

If external expertise is needed, contract the best available person in the world in the required field: do not compromise for less. Often, these first class people tend to be independent consultants with a deep knowledge in their field: people that wrote the book, not read it. Real consultants produce results immediately of far higher quality at lower cost; by comparison, pseudo-consultants from large firms have to investigate and study for weeks or months before producing any output, study time for which they expensively bill. For example, to create an XML software, development should be done by people with a deep XML knowledge, such as authors of `mod_xml` or `libxml2` [MXML][LXML]. Internal staff must absorb sufficient know-how to use components and not just passively rely on external consultants.

To make a bad situation worse, byzantine legalistic tender procedures must be respected, combined with the self-preserving instinct of individuals to avoid responsibilities; nobody was ever dismissed for following the *three bigs*: big consulting specification, big external development and big-bang. Some involved individuals count on being long gone to greener pastures by the big-bang day of reckoning.

Let's get real: even attempting to specify a comprehensive requirement for an organisation of a meaningful size is **futile** [CTO], a Gödel whiff [GOIT]; implementing it, is far more unrealistic. The best one can hope for is a frozen snap of a moving target, as the requirements are too large, complex, fuzzy, continuously changing; technology also moves at the same time. Attempting to prepare a comprehensive technical-legalistic tender for this kind of scenario is mirage grabbing. Throwing blindly people and money to problems does not work. But one cannot give up to desperation: something has to be done [BPW].

**New system**

{Principle} Think big, start small.

{Principle} *Bifocal*: address simultaneously short-term pressing needs and long-term strategy. *Sustainable system* is another view to this principle; paraphrasing Brundtland: *meeting present needs without compromising future needs*.

The guiding line is to evolve current systems focusing in fulfilling the most pressing needs in order of priority, without interrupting services or breaking anything.

Systems are needed today; but, by definition, **all systems are late**, as they should have been planned years before they are needed. Not only late, the vast majority of computer systems are over budget, and underperforming: the cost of poor software quality is horrendous [CPS].

One should be humble and readily admit that our planned new system will turn-out like the majority of computer systems and work hard to minimise these vices: keep a close eye on technology complexity and system functionalities, the main sources of risks and cost. Hence, one must use the simplest and lowest cost possible technology; introduce more complex and expensive technology when is really warranted. Restrict functionalities to what is immediately necessary and avoid too many functionalities, at least to start with. Aim for robustness, not fanciness.

Proven mature technologies are sufficient to cover the requirements of the vast majority of middle-of-the-road private or public organisations. Vanguard complex technologies are unnecessary. The approach proposed in this document is more than sufficient.

{Proverb} A bird in the hand is worth two in the bush.

Better to have a basic system today that provides 40% of the requirements than a promise of a wonderful system in four or five years. Basic systems can evolve by adding missing functionalities in less than four or five years, continuously adapting the specifications to the evolving requirements; the added bonus is that it provides some services immediately: less risky, and less traumatic [WPB][TPP].

One has to accept and formalise the reality without doctrinaire restrictions: properly maintaining existing systems that have been declared obsolete, but that fulfil current needs; building quickly evolutionary prototypes (not throwaway prototypes) to solve pressing needs. One has to keep in mind that the wonderful new system under development might not be so wonderful [PRO].

Having an incomplete computer system means accepting that uncovered needs will have to be done "manually"; for example, with infamous Excel spreadsheets [ES][RRE]. System evolution can be applied to enterprise system, but not necessarily to other fields: what about a plane with 40% completed avionics?

A good discipline is to estimate the real risks and costs of IT projects, which must be updated as changes arise, but the original estimation must always be remembered; hence, one is always confronted with the original estimations. As time passes, organisation amnesia kicks in on the original delivery date, functionalities and cost: accounting magic, project renaming and voilà a new project with all counters reset to zero; IThistory comes handy to highlight these magic tricks.

If using off-the-shelf systems, one has to add the integration costs with the current systems; this is often overlooked. Off-the-shelf systems that are an island and do not integrate become a headache; they become expensive and they take a disproportionate amount of resources. On the other hand, using off-the-shelf components (libraries, etc) are a very good approach. Note the difference of system vs. component.

Organisations must have sufficient IT resources to be the architect of their systems, though the majority of components could be subcontracted out; trying to avoid responsibilities and subcontracting out the whole system does not work and it is very expensive. This approach is for organisations of a significant size; small organisations are out of scope of this document and they might be better off hosting the whole lot externally as a service.

One has to avoid the vice of trying to create an internal IT firm: the organisation has a core business to attend to; computer systems are just a mean to an end, not the final objective. By the way, this type of vice is common to other supporting departments, such as: the training department pushing to create a university; translation department behaving as languages academies when the needs are plain translation, not literature translation; or security department sliding into transforming a building into a barrack.

The amount of time to get from a current typical system to a perpetual system varies widely: for *enterprises* as defined below, a very rough estimation is a range of **2 to 6 years**; this estimate is not for very large organisations such the National Gendarmerie in France [GB]. For a new enterprise starting a green field development, it should take just a few months: it is much easier as the migration cost and complexity are avoided.

**Documenting**
{Principle} Institutionalise knowledge: write down.

Before building a new system, it is essential describing the current organisation, not necessarily in an IT style: prose, perhaps a few diagrams. This is essential for whatever approach is followed: evolutionary or big-bang. It also has the additional benefit of clarifying for everybody the current functioning.

The guiding scenario is similar to documenting current IT systems: one must be able to find out answers to the functioning of any part of the organisation, though this ideal level of documenting might never be achieved.

Documenting often takes the form of updating and consolidating existing standard operating procedures (SOP), manuals, glossaries, better understanding of applicable legislations, etc. One might have strange surprises such as staff giving priority to manuals interpreting laws over the texts of laws. At the most basic level, the exact meaning of terms must be clear: terminological misunderstandings are common, frustrating and wasteful.

Undocumented parts have to be completed. Individuals usually have knowledge related to their work. Staff with deeper comprehensive understanding of the organisation is rarer. Often, staff disagree on how an organisation functions and they might have contradictory opinions. Also, there are human quirks that have to be addressed; for example, some staff like keeping their knowledge secret as it gives them a sense of power.

It is essential to have a chief editor responsible for the whole documentation, though individuals must contribute with their know-how, independently if it is in their field of responsibility or not. Collaborative writing tools are required, such as wikis. Documenting is a never ending task as organisations are continuously changing. The same principles must be applied to documenting IT systems.

**Final products**

Specifying final products helps to focus on what organisations have to deliver. For example, in the case of European institutions, these are mostly multilingual parallel documents, and the main category are legal acts [LA]. Having schemas [DTD] for legal acts would make clear what has to be produced; also they can be used as an interface (exchange format) among institutions. By the way, European legal acts are not formally specified.

It is essential to keep the eyes on the ball: the objective is final products; IT systems are just means to an end. IT systems do not have to be glittering, they have to be effective to deliver final products; as an illustration, what counts for customers are the cars, not the beauty of the factory.

**Human resources**

{Principle} Few outstanding internal IT staff.

{Definition} *Expertise cost*.- The cost of maintaining the required expertise to ensure the operation and maintenance of a computer system. Corollary 1: the more varied, complex, rare the required expertises, the higher the cost. Corollary 2: this is a fixed cost, independently if it is used or not.

The more general concept is *technology cost*, where expertise cost is one of the components. Technics must be the simplest possible widely known: finding this type of IT staff is easier and cheaper. Finding IT staff for a niche obscure technic is harder and more expensive. Complex technics, even if popular, require more qualified staff and hence it is more expensive. For example, developing with Ajax requires more expertise than simpler server-side technics, though these have less functionalities; hence, the use and additional cost of complex technics must be justified.

Expertise is a key factor: without it, systems are by definition legacy. From an organisational perspective, there are two expertise types: external and internal.

*External expertise* is needed to <u>develop</u> and maintain the *infrastructure components* such as operating system or database management system [DBMS]. One has to monitor that there are external entities curating these components and intervene, if necessary.

*Internal expertise* is needed to <u>use</u> infrastructure components and for developing and maintaining internal applications; for example, expertise to use Linux, not to contribute to the kernel development. For internal applications: project leadership and full responsibility in-house; design halfway with the assistance of the best experts in the world; coding outside as open source. The rationale is to minimise the need for internal expertise and hence staff numbers, though keeping in mind that certain things cannot be outsourced.

{Quote} *I am a programmer* - K. Thompson, Unix main designer [RTT].

IT staff must be <u>real</u> IT staff as opposed to the wave of <u>fake</u> IT staff with meaningless titles. A basic indicator is that IT staff must be programmers, even if not involved in programming or most of the programming is outsourced.

IT staff allocation within an organisation is an integral part for the evolution to a new system: a *distributed IT department* as opposed to a *centralised IT department*. Distributed IT departments contain a *horizontal infrastructure department* with its own IT staff to look after computer centres, networks, etc. IT staff working on applications should be distributed across the organisation closed to the people that know the business, though following the IT organisation rules so all the pieces fit together into one single system. Like pseudo-consultants, centralised IT departments are a serious danger: often they suck-in far more IT resources than they redistribute to the organisation.

**Legal**

A mine field: legal IT aspects must be addressed properly from the very beginning or they can become very expensive headaches. One needs legal certainty. The licenses of each item must be fully understood and followed, particularly the licenses related to the basic building blocks.

As much as possible, preference must be given to *free and open-source software* [FOSS]. With commercial licenses one must be extremely careful and really comprehend each little ramification: some of these licences are far harder to comprehend than the most twisted computer codes; *indirect access* [IA] is very illustrative.

Application programming interface (API) intellectual property rights has even deeper implications as it could affect basic building blocks such as POSIX [OG] and not even clean room design [CRD] is an escape. There is some protection following the decision of the U.S. Supreme Court in Google vs. Oracle [GVO].

**Enterprise**

*Enterprise* has the connotation of a <u>whole</u> organisation as opposed to just one department. The concept of enterprise is fuzzy. A rule of thumb: it should have one authority that can quickly impose decisions and not too big that it becomes an unstable monster.

{Principle} Specify interfaces with other enterprises.

Interfaces are required to cooperate with other enterprises. A parable is the Internet: each network is independent and they collaborate to form a whole. For example, one European institution or directorate-general (according to size) should be considered one enterprise that cooperates through interfaces with other European entities; do not try to make all the European institutions or a large institution one enterprise. The *Federal Enterprise Architecture* [FEA] should be a sobering example.

**Enterprise system**

*Enterprise system* has the connotation of integrated system expanding the whole organisation, for which is needed interoperability among the constituents subsystems. Integration is a continuum: from fully integrated subsystems to totally separated subsystems. Purpose built subsystems are fully integrated, while with legacy and third-parties systems one should expect a bumpier ride. There are human and machine integration aspects.

Humans perception of seamless integrated system is created by a consistent user interface. This requires consistent clients that could be: *graphical user interface* (GUI); web based using browsers; and *command line interface* [CLI], this mostly for programming and administration. Systems might implement these three client types, such in the case of email: Apple Mail, Google webmail and the command `mail` in Unix-like systems. Interface consistency breaks down when passing control to non-native systems; for example, passing control from a consistent system interface to a third-party such as Mediawiki.

Machines require interoperability. Integration smoothness depends on software interfaces refinements. For example, a purpose built subsystem might be realtime; legacy systems might rely on periodic data transfer.

There is a whole plethora of names for enterprise systems to choose from: enterprise resource planning (ERP), enterprise information system, enterprise software, enterprise application software, etc. Also according to the many different types, categories, views, tastes, etc: document management system, business intelligence, knowledge management, business activity monitoring, etc. New names appear as consulting firms *à la mode* publish new reports.

**Perpetual system**

{Definition} *Perpetual system*.- A computer system that lasts forever. Corollary 1: systems must be evolvable and scalable. Corollary 2: minimise dependency chains, ideally no external dependency [X2347][LP]. Corollary 3: understand the stack, particularly core technologies. Corollary 4: this needs more work.

Perpetual design is a general concept beyond IT. For example, the Alcántara Bridge [ALB] is nearly two millennia and as good as the first day. The term *perpetual* is inspired by the latin inscription on the bridge: *Pontem perpetui mansuram in saecula mundi* (perpetual bridge which will remain forever in the world). Even considering the different natures of stone bridges and computer systems, few IT designers would dare to state something similar even for a two decades system [HPO], as few IT systems are designed to be perpetual systems; indeed, some are perpetual by accident, not design.

Similarly to other fields, there should be an approach of *evidence-based IT*: use what has been proven to work. Existing long lasting systems should be studied even further for its perpetual proprieties; for example, the Internet, Linux or the computers in the Voyager program.

Perpetual systems might be viewed as parallel to long-term data preservation [DP]; a kind of long-term system preservation. Data has been much more studied, compared to systems.

**Perpetual enterprise system**
*Perpetual Enterprise System (PES)* is the application of perpetual system to enterprise system as presented in this document; it must be perceived as a <u>single seamless system</u> [P9]. Other *perpetual enterprise systems* (lower case) might be specified.

**PES architecture**
{Overstating the evident} Programming costs money (and angst).

{Principles}
- Copy: The nirvana of programming is not programming; if any programming is required, it must be done by the best people in the world in that domain. Copy also higher levels such as software patterns.

- Repurpose: for example, one can repurpose even old protocols such as *finger* [R742].

- *Running code wins* [RCW].

- Follow the accumulated wisdom from the *Unix philosophy* [UP][BUP][TAOUP], *The Elements of Programming Style* [EPS] and similar fountains of wisdom.

- Inbuilt monitoring and debugging: yes, computer systems have a tendency to fail and get attacked and this has to be detected very early.

- *The worst enemy of security is complexity* [APS][ACC].

- Zero trust architecture [ZTA].

- Autarkic: capable of functioning without relying in other systems, in particular disconnected from the Internet for services such as DNS.

- Full-stack: consider all parts, front-end, back-end and everything in the middle.

Before developing, one should look around to see if there is software that could be copied and repurposed for the needs at hand. Better still, the repurposing should be carried out externally by the designers of the original software: it will be by far the best work at the lowest cost. The resulting system must be put back in the public domain.

PES is open standard and preference is given to the use of FOSS. Perpetuity is reinforced by the use of well established artefacts such as the Internet technologies and Linux [IETF][W3C][ISO]. This is like betting on a horse that has already won the race.

Making internal projects FOSS have benefits: software have to be properly packaged and the main benefit is for internal use; projects are open to external criticisms; it avoids being forgotten, even if the projects are not commissioned. One can create open source projects and incorporate them into an existing organisation such as the Apache Software Foundation [ASF], hence contributing back to the community; care must be taken of not loosing intellectual property rights (IPR) and related aspects such as over-granting rights. The icing on the cake is that external organisations take care: less need for internal expertise and hence less cost, though there could be contributions; it increases software perpetuity.

Keeping projects secret makes sense when it gives competitive advantages to organisations. Even then, only part of the stack needs to be secret, and the commodity parts can be FOSS [BB].

In addition to fulfilling the internal needs, code must be <u>highly portable</u> and documented to facilitate the porting to the main operating systems: Linux (and Unix like), Microsoft Windows and Apple. Algorithms must be implemented as libraries with bindings to the main computer languages such as C, Bash, Java, JavaScript or Python; with the intention to facilitate the creation of server modules, graphical applications and CLI. Indeed, libraries might be implemented using a safer language such as Rust.

PES avoids using frameworks. For example, to build *containers*, it gives preference to the Linux natives *cgroups*, *chroot* and *namespaces*, as opposed to adopt whole frameworks such as Docker [DK]. Software stacks must be as thin as possible: using frameworks add fat and it makes internal systems dangerously more dependant on external infrastructure; Occam's razor. This can be viewed as minimising the *Software Bill of Materials* (SBOM).

Protocols, formats and similar components must be independent of the underlying operating system; for example, the email stack has been implemented in most operating systems.

Codes should be compiled from pristine sources for security reasons: to facilitate the *reproducible builds* [RB][RTT] and minimise supply chain attacks. The firmware should be coreboot [CB] or similar. Security for lower layers are out of scope: compiler, motherboard, processor, etc; though for more critical systems this might have to be included. Arguably, the biggest security danger is internal staff: developers and users. Publishing source codes increases security; better still, when combined with a bug bounty program.

Developers must have the perception of building a modular system where new requirements and technologies are easy to incorporate. As an illustration, it is like creating a new command for Linux.

Missing infrastructure peculiarities must be addressed at the appropriate level; for example, in the case of European institutions, intense multilinguality is very peculiar as it is rare to have organisations working in 24 languages, though modern software tends to be well internationalised. Other organisation types have their own peculiarities. Peculiarities must be part of the infrastructure; for example, before Unicode, character sets were a big expensive headache as each application tried badly to implement the needed character repertoire.

The software must be the simplest possible server-side code: HTML with CSS [HTML5][CSS]. JavaScript must be avoided and used only after a proper evaluation of the return on investment (ROI): it might be justifiable for external web sites that are the showrooms where products are sold, and for dynamic data visualisation [GM][D3][IB]. Static site generators are sufficient for certain use cases, faster for serving and safer for long-term archiving. The resulting output must work well at least on the major web browsers: Chrome, Firefox, Internet Explorer/Edge, Safari and Opera; also friendly to CLI programs such as `curl`, `wget` and `lynx`. In addition, it must also work with the PES Client.

If taking the more complex JavaScript approach, one should consider using *single-page application* (SPA), if possible avoiding the use of a JavaScript framework as this would make PES dependent on external components; building SPA from scratch is more costly, at least to start with. SPA reminds the Blit terminal.

PES has two parts: a *common part* and an *organisation part*. The common part is the same to all PES instantiations; it provides the environment, generic components and it encapsulates the most complex aspects. The organisation part is particular to each organisation; hence, there is a different organisation part per organisation, though large chunks might be reused for similar organisation types. There is no pretension to create a one size fits all enterprise system. This document addresses mainly the common part.

**Common Part**

The Common Part consists of: *PES Human Interface* (PEShuman); *PES Operating System* (PESos); *PES Network* (PESnet); *development components* such as programming languages, DBMS, and web server; *third-party systems* such as Mediawiki. All these are basic bricks to built PES.

Development components are used to build applications that follow PEShuman. Non-PES systems (legacy and third-party) with their own human interface breaks PEShuman; for example, running Mediawiki means that the human interface is dictated by Mediawiki and it will not be able to follow the PEShuman, though parametrisation might help.

**PES operating system (PESos)**

*PESos* is a bare minimal server operating system, just enough for hosting services, approach similar to Container Linux [CL] or Flatcar [FL]. It can be installed directly into the hardware, or virtualised particularly for development. Updating could either be whole, as in Container Linux, or with packages, though this would require the inclusion of a package manager. There must be a quick rollback mechanism to a previous stable version, if an update fails, such as self-check at boot time and there is some service failure, to automatically reboot with the previous sane version.

In practice today, it is a small Linux distribution, but it could be also implemented with other Unix-like operating systems such as Plan 9 [P9], or more generally POSIX. PESos could built from well established distributions such as Debian, without items not needed for operations such as X-like graphical user interface (GUI), compilers and manuals. A microkernel [L4] would remove even more fat.

Servers can be administered with a web interface; but also with CLI: time-tested, better for programatic administration with shellscripts and low cost.

By default, all ports are closed and no service runs, not even `https`, login process (`getty`) or `sshd`. The decision on which ports to open and which services to run is done at *Server mould* level; at least one service must run. There could be scenarios where all incoming ports are closed and a service initialises an outgoing connection. The rationale is to minimise the attack surface [CBS]. Additional security measures must be considered, such as root file system read-only.

**Server mould**

A server mould is a setup for specific use case; for example, a document archival service. There are many server moulds; the intention is to have a *Server Mould Collection* that would increase as new use cases come into light. A server mould could just be parametrisations or it might have additional software such as DBMS. Server moulds could be used *as is* or as a base to build organisation specific applications.

**PES Network (PESnet)**

*PESnet* is a network with the following main characteristics:

- Zeroconf.- The setup is fully automatic.

- Mesh.- Communication among nodes are peer-to-peer.

- Zero trust.- Clients and servers are responsible for their own security. The underlaying network is not trusted and all communications are encrypted. No protection is expected from firewalls or other mechanisms.

- Overlay.- It could be implemented as one or several overlay networks; for example, there could be an overlay network dedicated to the communication among servers.

Servers in PESnet run PESos. They are independent and follow the Unix principles of being *well-behaved* and *do one thing well*; also inspired by *self-contained system* [SCS] or *microservice* [MISE]. Servers host mostly stateless RESTful-like [REST] services; in general: one service, one server. Having separated services allows breaking complexity into digestible chunks, though one has to try to minimise the number of services.

Servers work in a *need to know basis* and each server knows its own dependencies; no need for a service discovery [SD]. The connections are peer-to-peer; among others, this minimises the needs for *load balancing* [LB].

*Mutual authentication* is strongly recommended among all participants, clients or servers, within PESnet or external. PESnet servers authentication is always mandatory. For PESnet clients, it is strongly recommended using *client certificates* [CC], so clients without the appropriate certificate can be excluded; every little helps. Users have their personal certificates as opposed to passwords, though the user certificate should be protected with a password. Some operations might be carried out without authentication; for example, web reading. Certificates are X.509. A step further would be adopting *multi-factor authentication* (MFA), for example with hardware security tokens such as YubiKey.

**PES profiles**

PES profiles are complete PES software packages to easily [X1742] install PES, either into one machine virtualising servers or distributed among several machines. Installation into one machine is intended for development; into several machines usually for production, though one machine could host several servers even in production. The capability of hosting PES in one machine is a good discipline: development, debugging, simulation, quality control, continuity plan, etc.

Note the difference between Server Mould (component) and PES Profile (complete system). As with a Server Mould Collection, the intention is to have a *PES Profile Collection* that would also increase as new use cases come into light and that could be used as is or as a base to build organisation specific PESs; just design patterns.

The PES profile *Hello World* [HW], a simple instantiation, that can be run and installed in a standalone computer disconnected from networks and where the PES servers are virtualised. This is useful in development, testing and demonstration.

Each PES instantiation has a name that could be the domain name. The *PES Dashboard* server provides a PES global view. Among others, it has the `pesinfo` that produces a comprehensive PES report, similar to `phpinfo` [PHPI]; web based and CLI.

Each PES instantiation for an organisation should have its own continuously updated PES profile. In addition to the first installation, it is essential to be able to quickly rebuild from scratch (programs and data separately) when it is not possible to restore from full backups, such as in cases of security breaches like SolarWinds. Indeed, there should be periodic exercises of installing from scratch the PES profile of the organisation in a test system.

One must aim for a fully self-managed system with automatic workflow without human intervention. One of the *fallacies of distributed computing* is that there is an administrator [FODC], where the best one can hope for is just a human supervisor.

**Client: Workstation**

End users communicate with PES mainly from web browsers; hence workstations must be able to run at least one of the major browsers such as Firefox; one has to <u>avoid the installation of binaries into the workstation</u>. PES does not rely on workstation operating systems (Windows, macOS, Linux); hence, PES uses only artefacts (such as formats and languages) supported natively by the browser without the need for external programs in the workstation operating system or add-ons to the browser.

*PES Client* is a dedicated PES web client (https), and it might support other protocols; this binary would have to be installed into each workstation. Server-side applications must work well with major browsers and the PES Client, though the PES Client will be more suited and more secure. Other clients could be implemented such as web interface for telephone screen and/or an app.

Having a light web based PES *Client* will improve the user experience, and in addition it will increase security. It can be built from scratch. An economical and fast way to create the PES Client is to hack a FOSS browser such as Firefox [FF] and remove all unnecessary features such as: add-ons, certificates authorities, buttons, URI line, JavaScript, etc. It must behave like portable applications that can run from a USB key and minimise any use of the underlying operating system; indeed, one could even envisage combining the client and security token into a USB. One can apply technics such as connection restricted to specific domains, certificates in the client and server, etc. Login could be based on certificates. The resulting client must be secure enough to be used in banking; i.e., banks must have confidence that they can build their own very secure clients from it.

An additional step would be to create a very thin minimum hardware *PES Terminal* dedicated to run the PES Client.

**URI**

{Definition} *Uniform Resource Identifier* (URI).- "a compact sequence of characters that identifies an abstract or physical resource" [STD66].

{Principle} From *everything is a file* to *everything is a URI* [EIEF].

URI is a way for naming things, a basic mechanism to build computer systems, particularly distributed systems. A *resource* can be anything, as opposed to just data. The definition of *resource* is kind of fuzzy, even in the standards:

- *Uniform Resource Identifier (URI): Generic Syntax* [STD66]: a resource is "... whatever might be identified by a URI ... resource is not necessarily accessible via the Internet; e.g., human beings, corporations, and bound books in a library can also be resources ..." .

- *Hypertext Transfer Protocol* [R2616]: "... network data object or service that can be identified by a URI ..." .

{Principle} One should think hard and decide once and for all on the URIs and particularly on the domain part.

Once minted, URIs must exist forever: *Cool URIs don't change* [CU]. URIs greatly help with *information hiding* [IH]; for example, URIs return documents independently of how they are stored: filesystem, document management system, static data, dynamic data, etc. URIs must be compact, though they should be mnemonic [MNE]; in particular, there must be no terminal strings such as `php` (the programming language in the server) or `html` (media type, format) except when requesting a certain *variant* [VAR], the media type must be in the HTTP header field `Content-Type`. A URI can have several variants; for example, the URI `http://example.com/rome` could have the Treaty of Rome in 30 languages, each in three formats, HTML, XML and PDF.

Read-only URIs are very efficient: read is cheap, write is expensive; more formally, using the GET method as opposed to the POST or PUT methods. It is a way of not moving data around.

Numbers are language neutral and unicase: they should be used in URIs, including in the host part; for example, `100.example.com`. This should have long-term stability as it is very compact, language and department independent. Numbers avoid problems with the URI *path section* that is case-dependant: lower and upper case characters are different; for example:

- these two are different `http://example.com/`**`foo`** `http://example.com/`**`FOO`**
- numbers are unicase `http://example.com/123`

**Data**

{Parable} PES is a self-modifying living organism, according to its own state and external stimuli from humans and machines. The things being modified is data.

PES is datacentric. The data catalog knows about all the data affecting PES: internal *PESdata* owned by PES; external data used by PES, for example weather service data. Some data drives programs logic that in turn modifies PESdata: logic/rules (business rules or others) must be soft-coded as modifiable data, as opposed to hardcoded into programs.

PESdata is distributed among the servers forming PES, where the data is kept in database, blockchains [BC] or other storage facilities. Each server must maintain its data dictionary, and all these data dictionaries have to be coordinated [DCAT][ACL]. All software must be database-agnostic, so different DBMS can be used. Preference is given to the simplest possible technology; for example: flat-file database, SQLite to more complex DBMS.

Reinstallation after an attack must take into account the data that drives programs as it is not a case of just reinstalling binaries. Unmutable data (append only) approach should be explored as data storage, disks and tapes are cheap.

**Data formats**
{Principle} Format the data, the how is secondary.

To be really useful, data must be structured, as opposed to amorphous. Format refers to data representation in files and protocols, as opposed to internal data structures of programs. Among other characteristics, formats must be:

- Human and machine friendly.
- Self-describing.
- Transformable into textual representations [TE].
- Processable without specialised programs; an editor should be sufficient.
- The finer the granularity, the better.
- Easy to marshalling/unmarshalling [MAR], between internal data structures of programs and external file/protocol formats.

Parsing is trivial with fine granularity. This spirit must permeate into other aspects such the HTTP message body [HMB], that should contain readily usable atomic pieces of data following the doctrine that *everything is an URI*, as opposed to fifty pages of Simple Object Access Protocol [SOAP] that have to be parsed to get to the relevant string. For example, when requesting the author of a work, the HTTP message body should contain only the string "`Joe Smith`", very readable for humans and machines.

Use the simplest possible format. For example, delimiter-separated values [DSV] is simpler than Extensible Markup Language [XML]. Isomorphic transformations (no loss of information) among formats are trivial. As with other IT technologies, pay the price for more complexity when it is really necessary; for example, XML is an overkill to represent tabular data, but it comes into its own when structuring documents.

Some relevant data formats/representations are: DSV for tabular data, key-value pairs [KV], record-jar [RJ], JSON [R7159], XML for documents, HTML the most web native [HTML5], Internet Message Format [R5322], SQLite [SL] for relational model. Fundamental data representations such as key-value pairs are implemented in several fashions [DF]. Preference must be for formats directly supported by the major browsers without the needed for external components, such as HTML, plain text, JSON, XML, SVG, JPG, PNG, GIF, etc.

The choice of formats comes down to a mixture of convenience, monies, and a bit of irrationality thrown in: *beware of the man of one book*. For example, XML is an awful format, but one has to be opportunistic as it is widely used for exchange (not necessarily for storage and processing) and there are many programs to process XML.

Data should be organised into packages that works standalone without a server and online in a server [XDOSSIER]. The package components can be of different formats.

**Data preservation**

{Principle} If one has two disks, one with the data and the other with the program, and only one can be saved from a fire, take the one with the data.

Data is paramount to PES, hence data preservation must have a very high priority: data must be <u>preserved forever</u>, so one is considering periods of 100 years and more [DP][R4810]. Addressing data preservation in depth is out of scope; what follows are a few hints.

Minimise the technology needed to access the data. Reading a proprietary binary format requires programs that in turn need their environments. Preferably, data should be <u>textual</u> [TE] and accessible from <u>filesystems</u>; for example, it must be easy to exchange (export/import) between DBMS and plain text. For rich presentation formats such as PDF, one must choose appropriate variations such as PDF/A [PA].

**Email**

{Principle} Email is <u>bad</u>: *email considered harmful* [CH].

Most organisations abuse the use of email, a horizontal ad hoc communication channel for functionalities hard to properly integrate into information systems. Processing and managing *routinary emails*, such as travel approvals, is inefficient and wasteful.

The existence of whole classes of routinary emails is a symptom of missing functionalities in PES. There must be a continuous email traffic analysis to identify routinary emails with the intention of eliminating them by adding PES missing functionalities. Usually, data in emails is amorphous, not structured data contributing to a datacentric PES.

One might use emails for notifications, if a proper notification functionality is missing. Notification emails might contain just a URI in the subject line and an empty body; they can be deleted, so no email management costs, and they must not contain data which must be in the concerned subsystem.

**Governance**

Principles and ideas in this document can be readily applied. The development of more technical parts of PES requires the creation of the PES Foundation or a project under an existing entity.

**Author**
Manuel Tomas CARRASCO BENITEZ
ca@dragoman.org
http://dragoman.org/pes

Little of the ideas presented are new, mostly rearranging and simplifying.